\documentclass[11pt,twocolumn]{article}
\usepackage{url}
\usepackage{hyperref}
\usepackage[cmex10]{amsmath}
\usepackage{amsfonts}
\usepackage{amssymb}
\usepackage{pgf}
\usepackage{graphicx}
\usepackage{grffile}
\usepackage{subfig}
\usepackage[font=scriptsize]{caption}
\usepackage{epstopdf}
\usepackage{mathtools}
\usepackage{tabu}
\usepackage{cite}
\usepackage[ruled,vlined]{algorithm2e}
\newcommand{\ud}{\,\mathrm{d}}
\usepackage[paper=a4paper,left=15mm,right=10mm,top=15mm,bottom=15mm]{geometry}
\usepackage{authblk}
\begin{document}
%


\title{Time-Frequency analysis via the Fourier Representation}
\author[1,2]{Pushpendra Singh \thanks{spushp@gmail.com; pushpendrasingh@iitkalumni.org}}
\affil[1]{Department of Electrical Engineering, IIT Delhi}
\affil[2]{Department of Electronics \& Communication Engineering, JIIT Noida}

\maketitle
\begin{abstract}
The nonstationary nature of signals and nonlinear systems require the time-frequency representation. In time-domain signal, frequency information is derived from the phase of the Gabor's analytic signal which is practically obtained by the inverse Fourier transform. This study presents time-frequency analysis by the Fourier transform which maps the time-domain signal into the frequency-domain.  In this study, we derive the time information from the phase of the frequency-domain signal and obtain the time-frequency representation. In order to obtain the time information in Fourier domain, we define the concept of `frequentaneous time' which is frequency derivative of phase. This is very similar to the group delay, which is also defined as frequency derivative of phase and it provide physical meaning only when it is positive.
The frequentaneous time is always positive or negative depending upon whether signal is defined for only positive or negative times, respectively. If a signal is defined for both positive and negative times, then we divide the signal into two parts, signal for positive times and signal for negative times. The proposed frequentaneous time and Fourier transform based time-frequency distribution contains only those frequencies which are present in the Fourier spectrum.
Simulations and numerical results, on many simulated as well as read data, demonstrate the efficacy of the proposed method for the time-frequency analysis of a signal.


\end{abstract}
\section*{INTRODUCTION}

The time-domain representation and the frequency-domain representation are two classical representations of a signal. In
both domains, the time ($t$) and frequency ($f$) variables are mutually exclusive. The time-frequency distribution (TFD) on the other hand, provides localized signal information in time and frequency domain. The TFD provides insight into the complex structure of a signal consisting of
several components. There exist many types of time-frequency analysis methods such as short-time Fourier transform, Gabor transform, Wavelet transforms, and Wigner-Ville distribution.

The Carson and Fry (1937) introduced~\cite{th19} the concept of variable frequency, with application to the theory of frequency modulation (FM), as a generalization of the definition of constant frequency. Moreover, the nonstationary nature of the signals and nonlinear systems require the idea of instantaneous frequency (IF). The IF is the basis of the TFD or time-frequency-energy (TFE) analysis of a signal. The IF is a practically important parameter of a signal which can reveal the underlying process and provides explanations for physical phenomenon in many applications such as vibration, acoustic, speech signal analysis~\cite{rslc9}, meteorological and atmospheric applications~\cite{rslc1}, seismic~\cite{rslc9}, radar, sonar, solar physics, structural engineering, communications, health monitoring, biomedical and medical applications~\cite{th46}, cosmological gravity wave and financial market data analysis.

The IF is the time derivative of the instantaneous phase that is obtained by the Gabor's complex signal, which is well-known as the analytic signal, representation~\cite{Gabor,th19,th20,th21,th22,th23}. The IF is well-defined only when time derivative of phase is positive, if this derivative is negative then IF does not provide any physical significance~\cite{rslc1,rslc2,rslc3,rslc4,rslc6,rslc71,rslc72,rslc73,rslc8,rslc9,th4,th411}. In order to remove this problem, recently many nonlinear and nonstationary signal representation, decomposition and analysis methods, e.g. empirical mode decomposition (EMD) algorithms~\cite{rslc1,rslc2,rslc3,rslc4,rslc5,rslc6,rslc7}, synchrosqueezed wavelet transforms (SSWT)~\cite{rslc71}, variational mode decomposition (VMD)~\cite{rslc72}, eigenvalue decomposition (EVD)~\cite{rslc73} and Fourier decomposition methods (FDM)~\cite{rslc8,rslc9,rslc10,rslc11,rslc12,rslc13}, are proposed. The main objective of all these methods is to obtain the representation or decomposition of a signal such that the IF is positive for all time.

Recently, Singh~\cite{psinghBL} has obtained an important enhancement in the definition of the IF where it is redefined in such that it is valid for all types of signals such as monocomponent and multicomponent, narrowband and wideband, stationary and nonstationary, linear and nonlinear signals. This has been obtained by redefining the IF such that it is always positive by using the fact that inverse tangent is a \emph{multivalued} (i.e. one-to-many mapping) function. Thus, this definition of IF has provided a way to obtain TFD of a signal by decomposing into a set of desired frequency bands.

As is well-known that the Fourier theory, which maps a time-domain signal into the frequency-domain signal, is the only tool for spectrum analysis of a signal. The FDM~\cite{rslc8,rslc9} has demonstrated that it is also a superior tool for nonlinear and nonstationary time series analysis. In this study, we demonstrate that the time information, which we refer to as `frequentaneous time' as a dual of the `instantaneous frequency', can be recovered from the frequency derivative of phase of frequency domain signal. So obtained frequentaneous time $\tau(f)$, frequency $f$ and Fourier domain signal $X(f)$ are used to obtain the three dimensional time-frequency distribution of a signal.



\section*{METHODS}
The Fourier series is a well-known and most important representation of a periodic function in the mathematics, science and engineering for spectral analysis of a physical phenomena.
The discrete time Fourier transform (DTFT) of a signal $x[n]$ is defined as
\begin{equation}
X(\omega)= \sum_{n=-\infty}^{\infty} x[n] e^{-j\omega n} =X_r(\omega)+jX_i(\omega), \label{Ch1_eq1}
\end{equation}
where $X_r(\omega)$ and $X_i(\omega)$ are real and imaginary part of $X(\omega)$, respectively. The original signal $x[n]$ can be obtained from $X(\omega)$ by the inverse DTFT (IDTFT), which is defined as
\begin{equation}
x[n]= \frac{1}{2\pi}\int_{-\pi}^{\pi} X(\omega) e^{j\omega n} \ud \omega. \label{Ch1_eq2}
\end{equation}
One can easily observe that in \eqref{Ch1_eq1} signal is being averaged over time to obtain frequency domain signal, and in \eqref{Ch1_eq2} signal is being averaged over frequency to obtain time domain signal.

Now onwards we always assume, unless otherwise specified, that $x[n]$ is a real-valued function, then $X_r(\omega)$ is a even function, $X_i(\omega)$ is odd function (i.e $X_r(\omega)=X_r(-\omega)$ and $X_i(\omega)=-X_i(-\omega)$) and phase spectrum is odd function of frequency (i.e. $\phi(\omega)=-\phi(-\omega)$ where $\phi(\omega)=\tan^{-1}[X_i(\omega)/X_r(\omega)]$). Thus, real and imaginary part are always orthogonal, i.e.
\begin{equation}
\int_{-\pi}^{\pi} X_r(\omega)X_i(\omega) \ud \omega=0,\label{Ch1_eq1_0}
\end{equation}
because multiplication of even and odd function is odd function and integration of odd function in a limit $[-a, a]$ is always zero.

As $x[n]$ is a real-valued function, from \eqref{Ch1_eq2} we can obtain analytic signal (Fourier transform vanishes for negative frequencies)
\begin{equation}
z_1[n]= \frac{1}{\pi}\int_{0}^{\pi} X(\omega) e^{j\omega n} \ud \omega = z_{1r}[n]+jz_{1i}[n], \label{Ch1_eq3}
\end{equation}
and dual-analytic signal (Fourier transform vanishes for positive frequencies)
\begin{equation}
\tilde{z}_1[n]= \frac{1}{\pi}\int_{-\pi}^{0} X(\omega) e^{j\omega n} \ud \omega = \tilde{z}_{1r}[n]+j\tilde{z}_{1i}[n], \label{Ch1_eq4}
\end{equation}
such that $x[n]={\left(z_1[n]+\tilde{z}_1[n]\right)}/{2}$ and $z^*_1[n]=\tilde{z}_1[n]$, where $*$ denotes complex conjugate operation.

An analytic signal $z_1[n]$ can be written as
\begin{equation}
 \left.\begin{aligned}
        z_1[n] & =z_{1r}[n]+jz_{1i}[n] = a_1[n]e^{j\phi_1[n]},\\
    \text{ where }    a_1[n] & =\left[z^2_{1r}[n]+z^2_{1i}[n]\right]^{1/2}\ge 0, \\
    \text{ and }    \phi_1[n] & =\tan^{-1}\left(z_{1i}[n]/z_{1r}[n]\right).
       \end{aligned}
 \right\} \label{Ch1_eq5}
\end{equation}

The IF for this analytic signal is defined in~\cite{psinghBL} as
\begin{equation}
 \omega_1[n] =
  \begin{cases}
    \big(\phi_1[n] -\phi_1[n-1] \big) & \text{ if difference is} \geq 0, \\
   \big(\phi_1[n] -\phi_1[n-1] \big)+\pi & \text{ otherwise,}
  \end{cases}\label{Ch1_eq6}
\end{equation}
where $\phi_1[n]$ is unwrapped phase, because phase unwrapping is necessary to ensure that all appropriate multiples of $2\pi$ have been included in phase angle. Phase unwrap operation corrects the radian phase angles by adding multiples of $\pm 2\pi$ when absolute jumps between consecutive elements of a phase vector are greater than or equal to the default jump tolerance of $\pi$ radians~\cite{matlabweb}.

One can obtain the TFE distribution of a signal by 3-D plot of $\{n, \omega_1[n], a^2_1[n]\}$.
From this 3D TFE distribution, we sum over the frequency (integrate over frequency in case of continuous signal) and obtain $a^2_1[n]$ which is true instantaneous energy (i.e. square of amplitude).

Similarly, a dual-analytic signal $\tilde{z}_1[n]$ can be written as
\begin{equation}
 \left.\begin{aligned}
        \tilde{z}_1[n] & =\tilde{z}_{1r}[n]+j\tilde{z}_{1i}[n] = \tilde{a}_1[n]e^{j\tilde{\phi}_1[n]},\\
    \text{ where }    \tilde{a}_1[n] & =\left[\tilde{z}^2_{1r}[n]+\tilde{z}^2_{1i}[n]\right]^{1/2}\ge 0, \\
    \text{ and }    \tilde{\phi}_1[n] & =\tan^{-1}\left(\tilde{z}_{1i}[n]/\tilde{z}_{1r}[n]\right).
       \end{aligned}
 \right\} \label{Ch1_eq7}
\end{equation}
The IF, which is always negative, for this dual-analytic signal we define as
\begin{equation}
 \tilde{\omega}_1[n] =
  \begin{cases}
    \big(\tilde{\phi}_1[n] -\tilde{\phi}_1[n-1] \big) & \text{ if difference is} \leq 0, \\
   \big(\tilde{\phi}_1[n] -\tilde{\phi}_1[n-1] \big)-\pi & \text{ otherwise.}
  \end{cases}\label{Ch1_eq8}
\end{equation}
It is to be noted that in the analytic signal (time-domain signal), frequency information is encoded in phase. In a dual way, one can observe that in the frequency-domain signal, time information is encoded in phase.

In order to decompose the analytic signal into a set of desired orthogonal frequency bands, we write~\eqref{Ch1_eq3} as
\begin{equation}
z_1[n]=\frac{1}{\pi}\int_{0}^{\pi} X(\omega) e^{j\omega n} \ud \omega=\sum_{i=1}^M a_i[n]e^{j\phi_i[n]} \label{FDM_eq25}
\end{equation}
where (with $\omega_0=0, \omega_M=\pi$)
\begin{equation}
 a_i[n]e^{j\phi_i[n]} =  \frac{1}{\pi}\int_{\omega_{i-1}}^{\omega_i} X(\omega) e^{j\omega n} \ud \omega, \label{FDM_eq26}
\end{equation}
for $i=1,\cdots,M$.

Now, from \eqref{Ch1_eq1} we can obtain
\begin{equation}
X_1(\omega)= \sum_{n=0}^{\infty} x[n] e^{-j\omega n} =X_{1r}(\omega) +j X_{1i}(\omega)\label{Ch1_eq9}
\end{equation}
and
\begin{equation}
\tilde{X}_1(\omega)= \sum_{n=-\infty}^{-1} x[n] e^{-j\omega n}=\tilde{X}_{1r}(\omega) +j \tilde{X}_{1i}(\omega) \label{Ch1_eq10}
\end{equation}
such that $X(\omega)=X_1(\omega)+\tilde{X}_1(\omega)$.

The signal $X_1(\omega)$ can be written as
\begin{equation}
 \left.\begin{aligned}
        X_1(\omega) & =X_{1r}(\omega) +j X_{1i}(\omega) = a_1(\omega)e^{j\phi_1(\omega)},\\
    \text{ where }    a_1(\omega) & =\left[X^2_{1r}(\omega)+X^2_{1i}(\omega)\right]^{1/2}\ge 0, \\
    \text{ and }    \phi_1(\omega) & =\tan^{-1}\left[X_{1i}(\omega)/X_{1r}(\omega)\right].
       \end{aligned}
 \right\} \label{Ch1_eq11}
\end{equation}
Here, we define frequentaneous time (frequency derivative of phase) for this signal as
\begin{equation}
\tau_1(\omega)=-\frac{\ud}{\ud f} \phi_1(\omega). \label{Ch1_eq12}
\end{equation}
It is to be noted that the frequentaneous time is even function of frequency because phase spectrum is odd function of frequency and differentiation of odd function is always even function.

In order to define frequentaneous time which is always positive for signal $X_1(\omega)$, we consider discrete version of this, $X_1[k]$, and define
\begin{equation}
 \tau_1[k] =
  \begin{cases}
    -\big(\phi_1[k] -\phi_1[k-1] \big) & \text{ if difference is} \geq 0, \\
   -\big(\phi_1[k] -\phi_1[k-1] \big)+\pi & \text{ otherwise,}
  \end{cases}\label{Ch1_eq13}
\end{equation}
where $\phi_1[k]$ is unwrapped phase.

Similarly, we write signal $\tilde{X}_1(\omega)$ as
 \begin{equation}
 \left.\begin{aligned}
        \tilde{X}_1(\omega) & =\tilde{X}_{1r}(\omega) +j \tilde{X}_{1i}(\omega) = \tilde{a}_1(\omega)e^{j\tilde{\phi}_1(\omega)},\\
    \text{ where }    \tilde{a}_1(\omega) & =\left[\tilde{X}^2_{1r}(\omega)+\tilde{X}^2_{1i}(\omega)\right]^{1/2}\ge 0, \\
    \text{ and }    \tilde{\phi}_1(\omega) & =\tan^{-1}\left(\tilde{X}_{1i}(\omega)/\tilde{X}_{1r}(\omega)\right),
       \end{aligned}
 \right\} \label{Ch1_eq14}
\end{equation}
and we define frequentaneous time which is always negative for signal $\tilde{X}_1(\omega)$ by considering discrete version of this, $\tilde{X}_1[k]$, as
\begin{equation}
 \tilde{\tau}_1[k] =
  \begin{cases}
    -\big(\tilde{\phi}_1[k] -\tilde{\phi}_1[k-1] \big) & \text{ if difference is} < 0, \\
   -\big(\tilde{\phi}_1[k] -\tilde{\phi}_1[k-1] \big)-\pi & \text{ otherwise,}
  \end{cases}\label{Ch1_eq15}
\end{equation}
where $\tilde{\phi}_1[k]$ is unwrapped phase.
Here, we obtain the TFE distribution of a signal by 3-D plot of $\{k, \tau_1[k], a^2_1[k]\}$ and $\{k, \tilde{\tau}_1[k], \tilde{a}^2_1[k]\}$.
From this 3D TFE distribution, if we sum over the time (integrate over time in case of continuous signal) then we obtain marginal spectrum $a^2_1[k]$ and $\tilde{a}^2_1[k]$, which is true Fourier based power spectral density (PSD).

\textbf{Discussion:} Notice that the both TFE distribution (1) obtained by~\eqref{Ch1_eq5} and ~\eqref{Ch1_eq6} (i.e. with $\{n, \omega_1[n], a^2_1[n]\}$ and it is well-known as Hilbert spectrum), and (2) obtained by this proposed method (i.e. with $\{k, \tau_1[k], a^2_1[k]\}$ and $\{k, \tilde{\tau}_1[k], \tilde{a}^2_1[k]\}$) are using the Fourier theory only. In order to obtain the TFE distribution of a signal, first method is using the inverse Fourier transform to obtain the analytic representation and second one (i.e. this proposed method) is using the forward Fourier transform. Thus, practically, both theses methods only rely on the Fourier theory directly. Hence, here in this study, we refer first method as Fourier-Hilbert spectrum (FHS) or Time-Frequency Distribution (TFD) by `instantaneous frequency' (TFD-IF) and second one as TFD by `frequentaneous time' (TFD-FT).

In all the above discussions, we have considered discrete time (DT) signal processing using the DTFT and DFT, which can be very easily generalized for continuous time (CT) signal processing using CT Fourier transform (CT-FT) and CT Fourier series (CT-FS).

%

\section*{RESULTS AND DISCUSSION}
In this section, we consider number of examples that are mostly discussed in literature to validate the efficacy of method under study.

\begin{figure}[!t]
\centering
\includegraphics[angle=0,width=0.5\textwidth,height=0.3\textwidth]{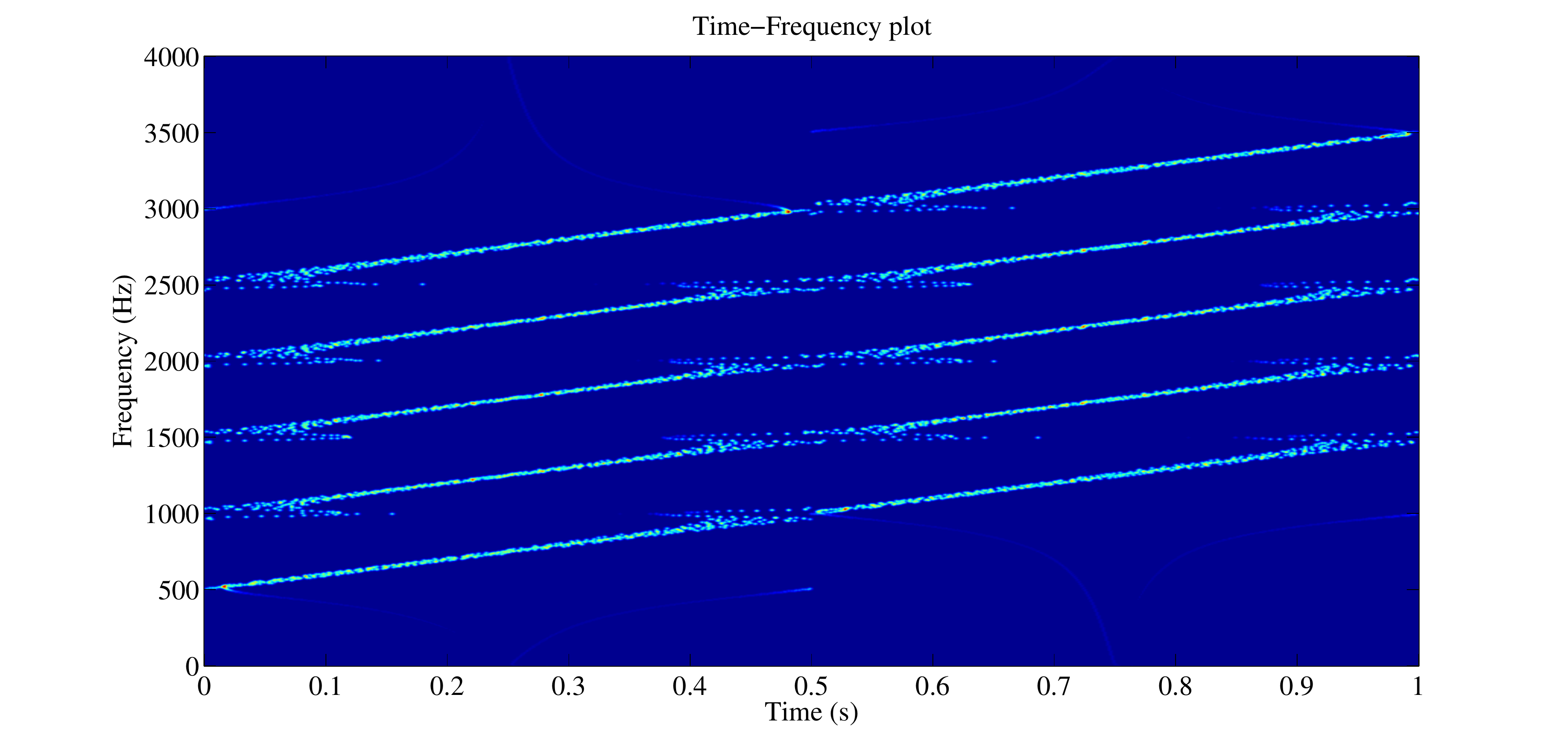}
\includegraphics[angle=0,width=0.5\textwidth,height=0.3\textwidth]{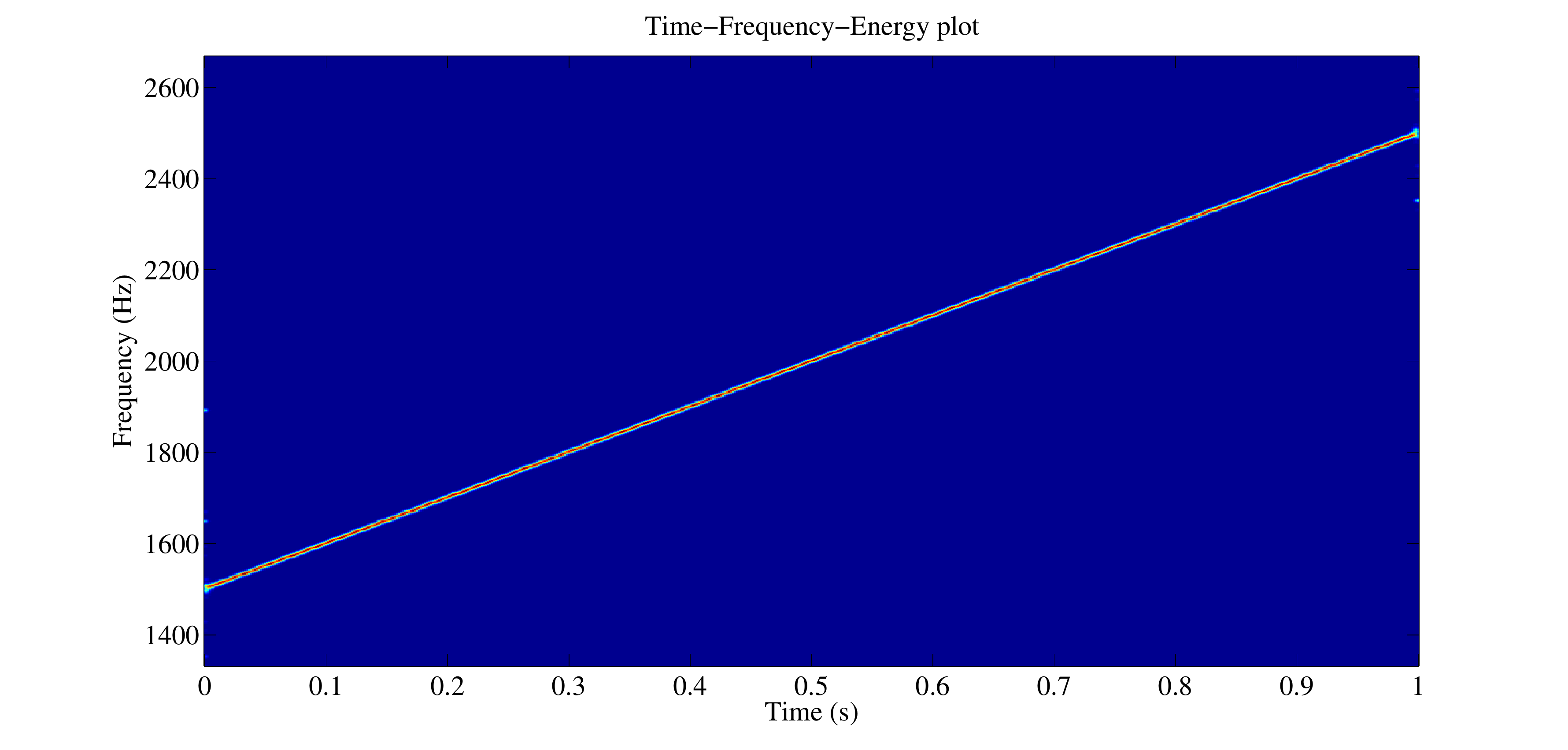}
\includegraphics[angle=0,width=0.5\textwidth,height=0.3\textwidth]{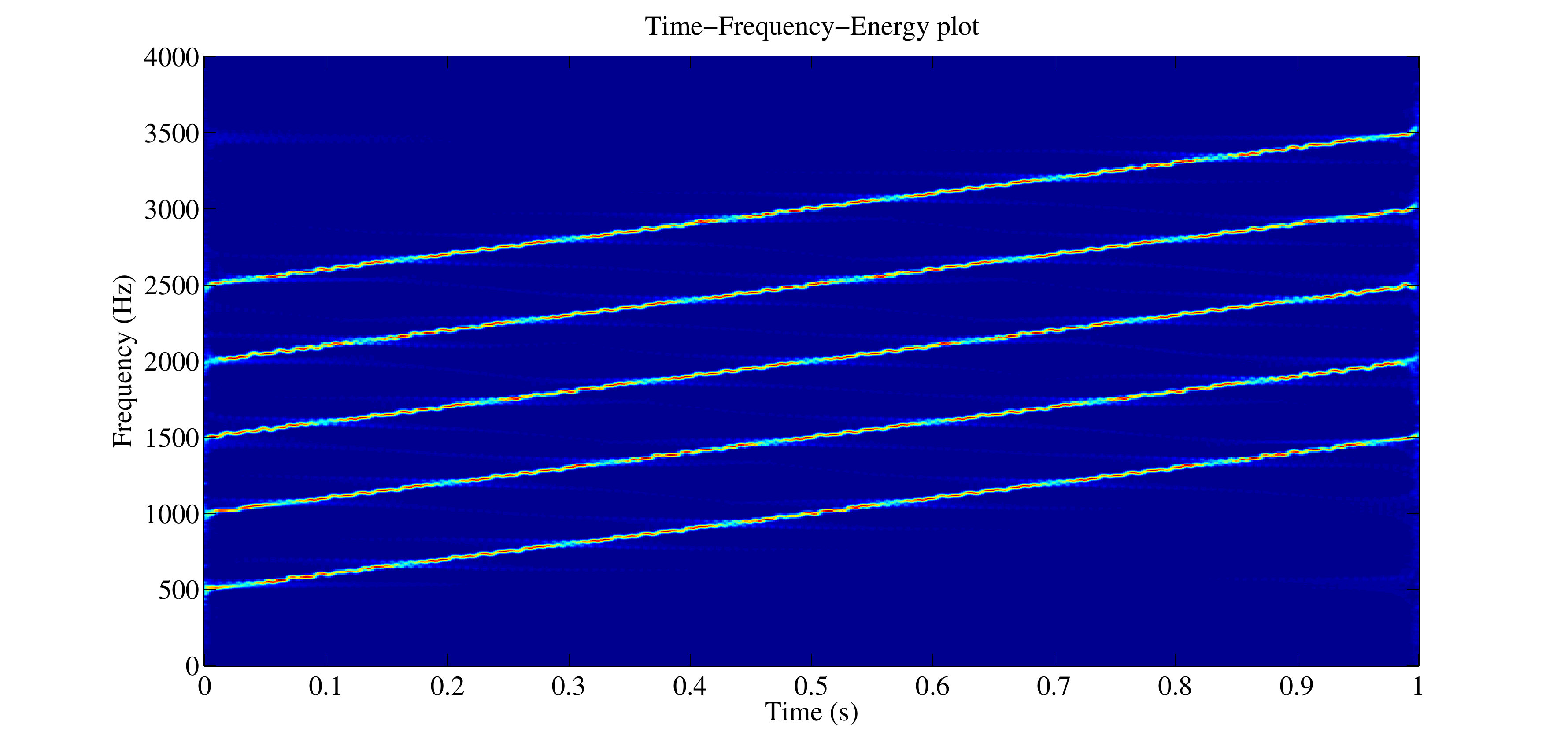}
\captionof{figure}{The TFE analysis of a nonstationary signal, which is sum of five linear chirp signals, using  TFD-FT (top figure), using TFD-IF (middle) without decomposition, (bottom) with decomposition into 30 bands of equal frequencies.}
\label{fig:ParallelChirps}
\end{figure}

\textbf{Example 1:} We obtain a nonstationary signal by adding five linear chirps of frequencies [500--1500] Hz, [1000--2000] Hz, [1500--2500] Hz, [2000--3000] Hz and [2500--3500] Hz. Figure~\ref{fig:ParallelChirps} shows the TFE analysis of this nonstationary signal, which is sum of five linear chirp signals, using proposed method TFD-FT (top figure), using TFD-IF (middle figure) without decomposition that presents average frequencies [1500--2500] Hz, which are average of frequencies present in five chirp signals; (bottom figure) with decomposition into 30 bands of equal frequencies. These two (top and bottom one) figures clearly reveal the five chirp signals present in the signal under analysis.

\begin{figure}[!t]
\centering
\includegraphics[angle=0,width=0.5\textwidth,height=0.3\textwidth]{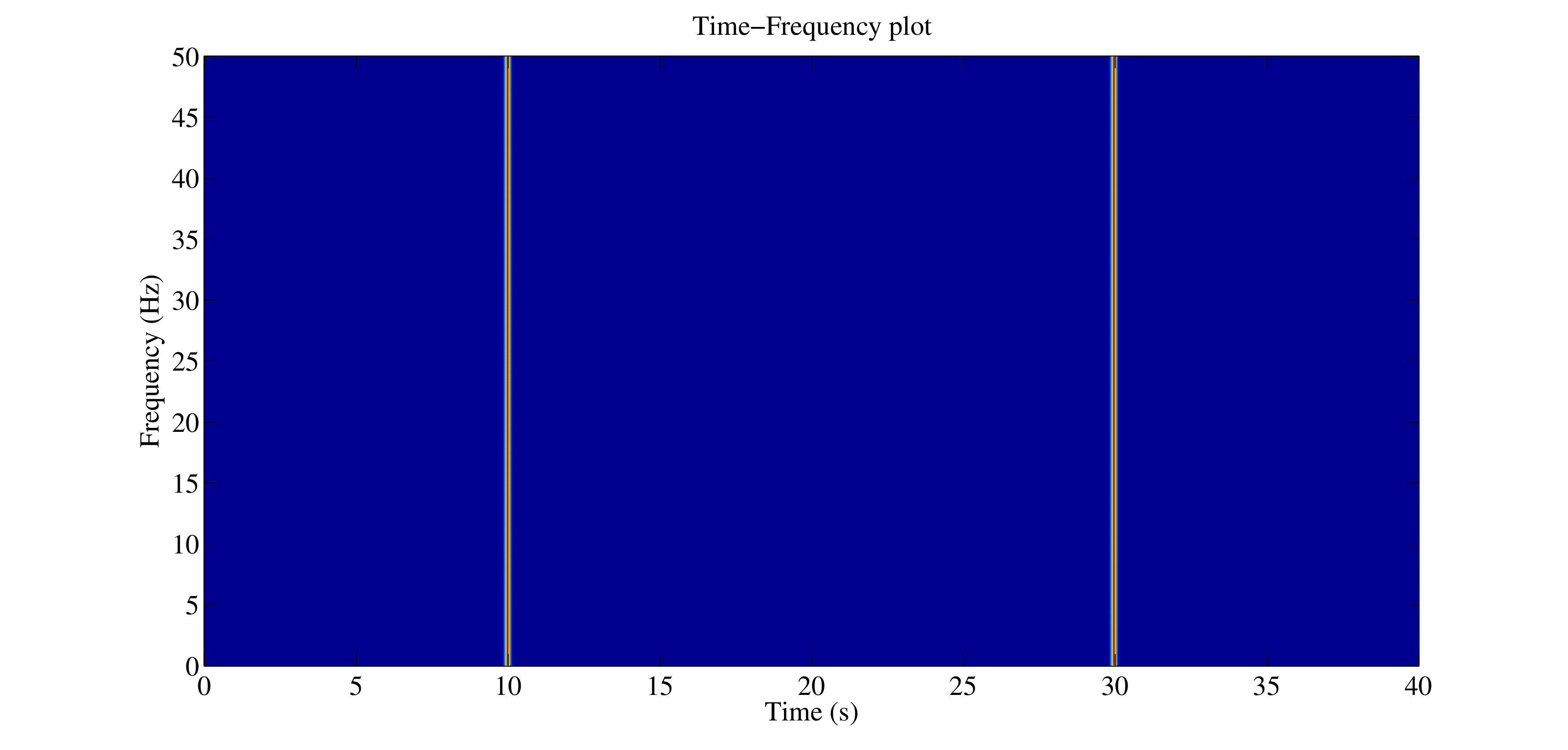}
\includegraphics[angle=0,width=0.5\textwidth,height=0.3\textwidth]{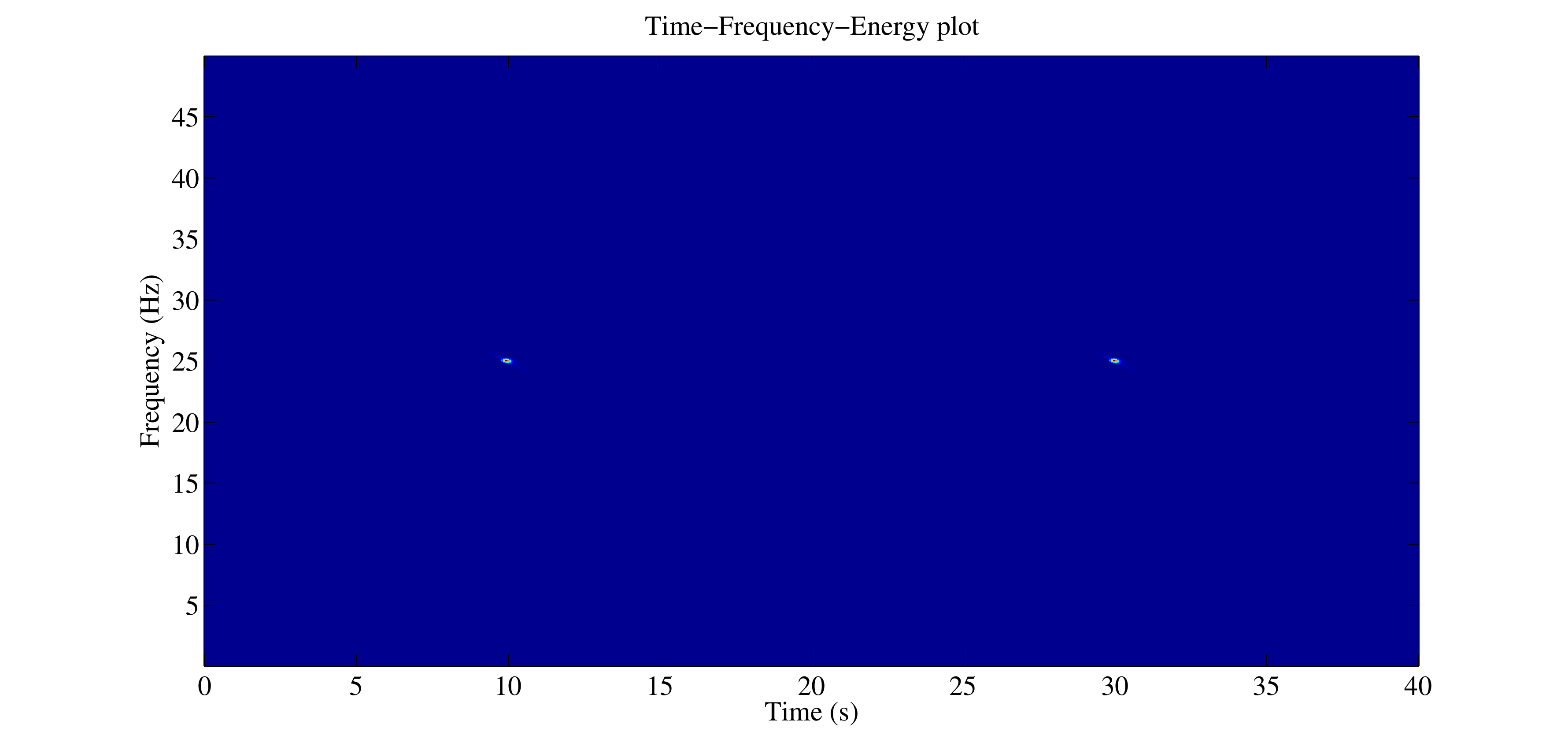}
\includegraphics[angle=0,width=0.5\textwidth,height=0.3\textwidth]{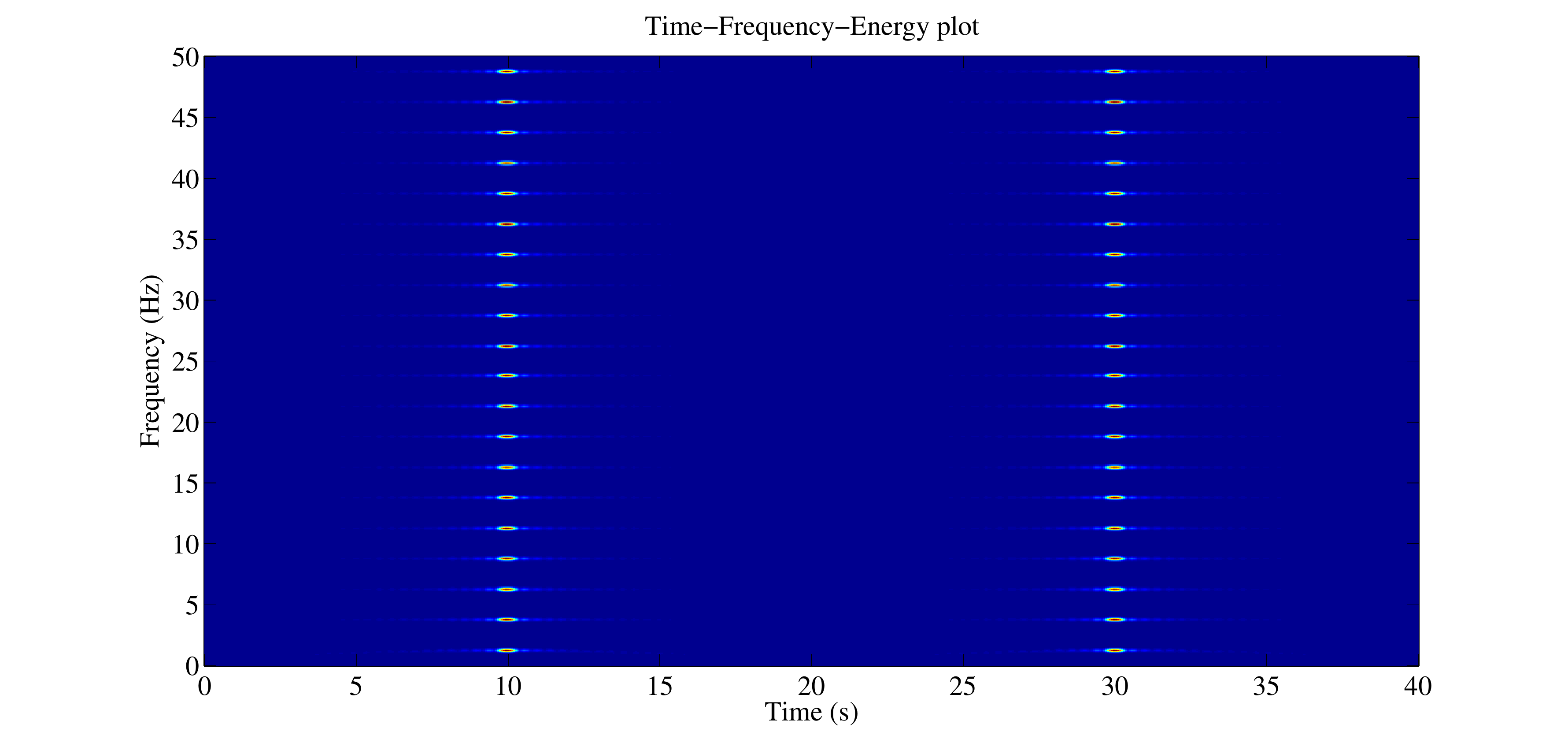}
\captionof{figure}{The TFE analysis of sum of two unit sample sequence $\delta[n-n_0]$ (with, $n_0 = 1000, n_1=3000$,
sampling frequency Fs = 100 Hz, length $N = 4000$) using TFD-FT (top), TFD-IF (middle) without decomposition and TFD-IF (bottom) with decomposition into 20 bands of equal frequencies.}
\label{fig:ui}
\end{figure}

\textbf{Example 2:} Here, we consider a signal which is sum of two unit sample sequences (delta functions) defined as $x[n]=\delta [n-n_0]+\delta [n-n_1]$ with $n_0=1000$ and $n_1=3000$.
It is well-known that delta function is a superposition of equal amplitude sinusoidal functions of all frequencies [0--$\pi$). The Nyquist frequency ($F_s/2$) is the highest frequency that can be present at a given sampling rate, $F_s$, in a discrete-time signal.
Figure~\ref{fig:ui} shows the TFE estimates of this signal (with $Fs = 100$ Hz and length $N = 4000$) using the TFD-FT (top figure),  TFD-IF (middle figure) without decomposition and TFD-IF (bottom figure) with decomposition into 20 bands of equal frequencies.
We observe that the frequency present in TFD-FT plot are true frequencies and signal is concentrated on time and spread over all the frequencies (i.e. it follows the uncertainty principle). However, TFD-IF plot without decomposition provide a average frequency (because delta function contains equal amplitude sinusoids of all frequencies form 0 to $F_s/2$) plot where signal is concentrated in time-frequency plane (i.e. it may not follow the uncertainty principle but frequencies present in plane are average frequencies and not true frequencies), and TFD-IF plot with decomposition provide a average frequency in decomposed bands.

\begin{figure}[!t]
\centering
\includegraphics[angle=0,width=0.5\textwidth,height=0.3\textwidth]{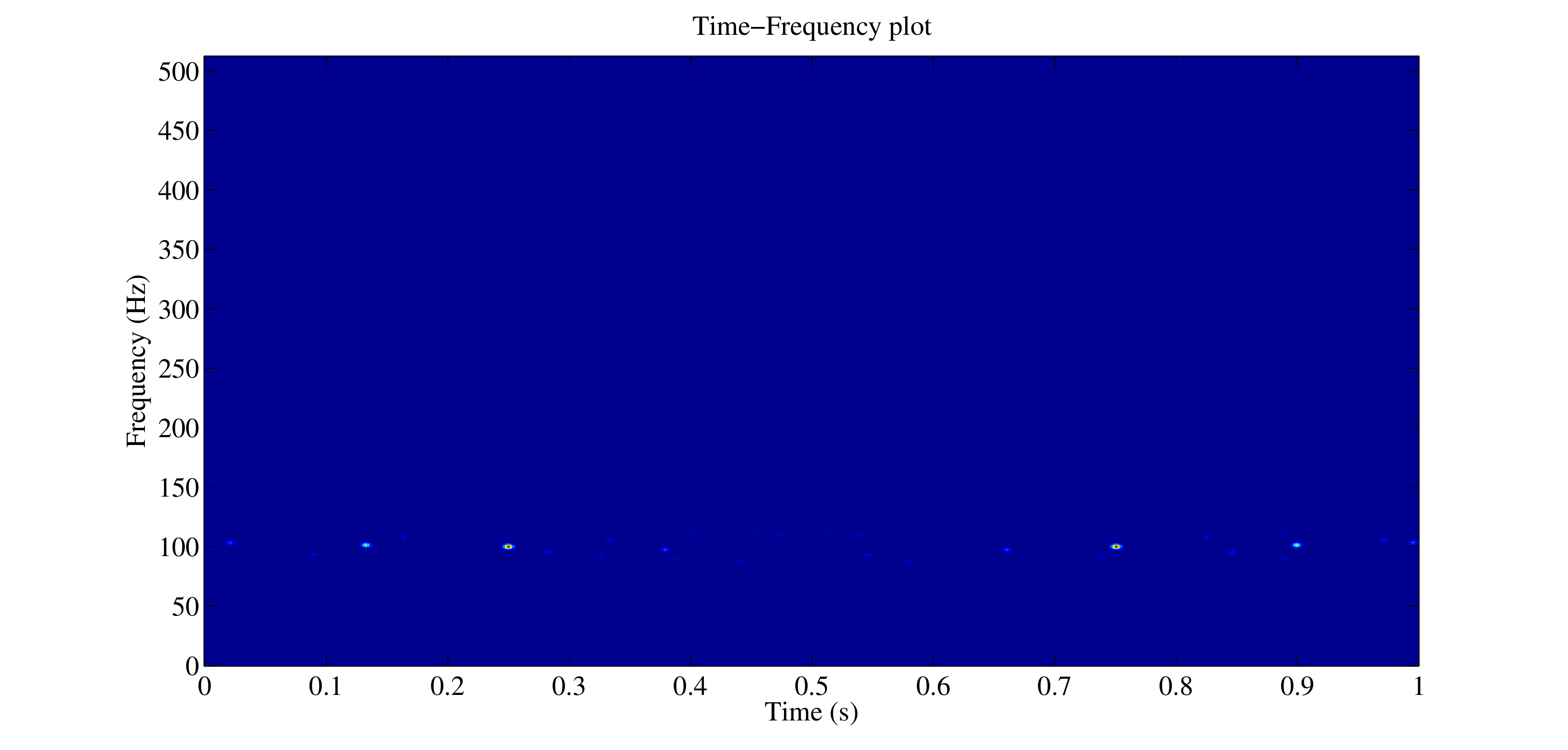}
\includegraphics[angle=0,width=0.5\textwidth,height=0.3\textwidth]{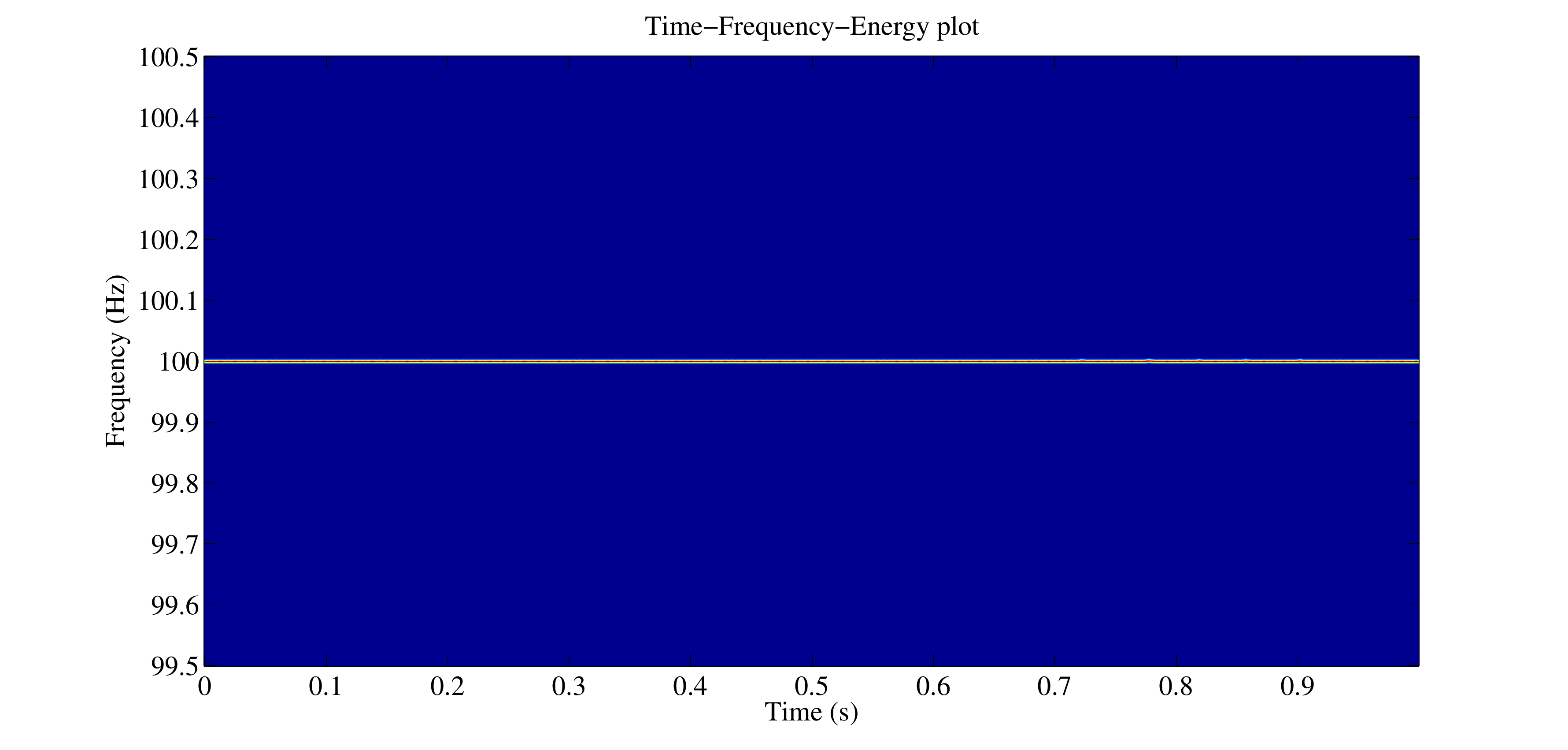}
\captionof{figure}{The TFE analysis of a sinusoidal function $f=100$ Hz (with sampling frequency Fs = 1000 Hz, length $N = 1000$) using TFD-FT (upper) and TFD-IF (lower).}
\label{fig:single_freq}
\end{figure}

\textbf{Example 3:} In this example, we consider sinusoidal signals which are concentrated in frequency and spread over the time. Figure~\ref{fig:single_freq} shows the TFE analysis of a sinusoidal function of $f=100$ Hz frequency (with sampling frequency Fs = 1000 Hz, length $N = 1000$) using TFD-FT (upper) and TFD-IF (lower). Due to time averaging in the TFD-FT (upper) plot, it shows the signal under analysis is concentrated in one frequency ($100$ Hz) and some time instants, whereas TFD-IF (lower) plot provides correct representation where signal is concentrated in frequency and spread over all the time.

\begin{figure}[!t]
\centering
\includegraphics[angle=0,width=0.5\textwidth,height=0.3\textwidth]{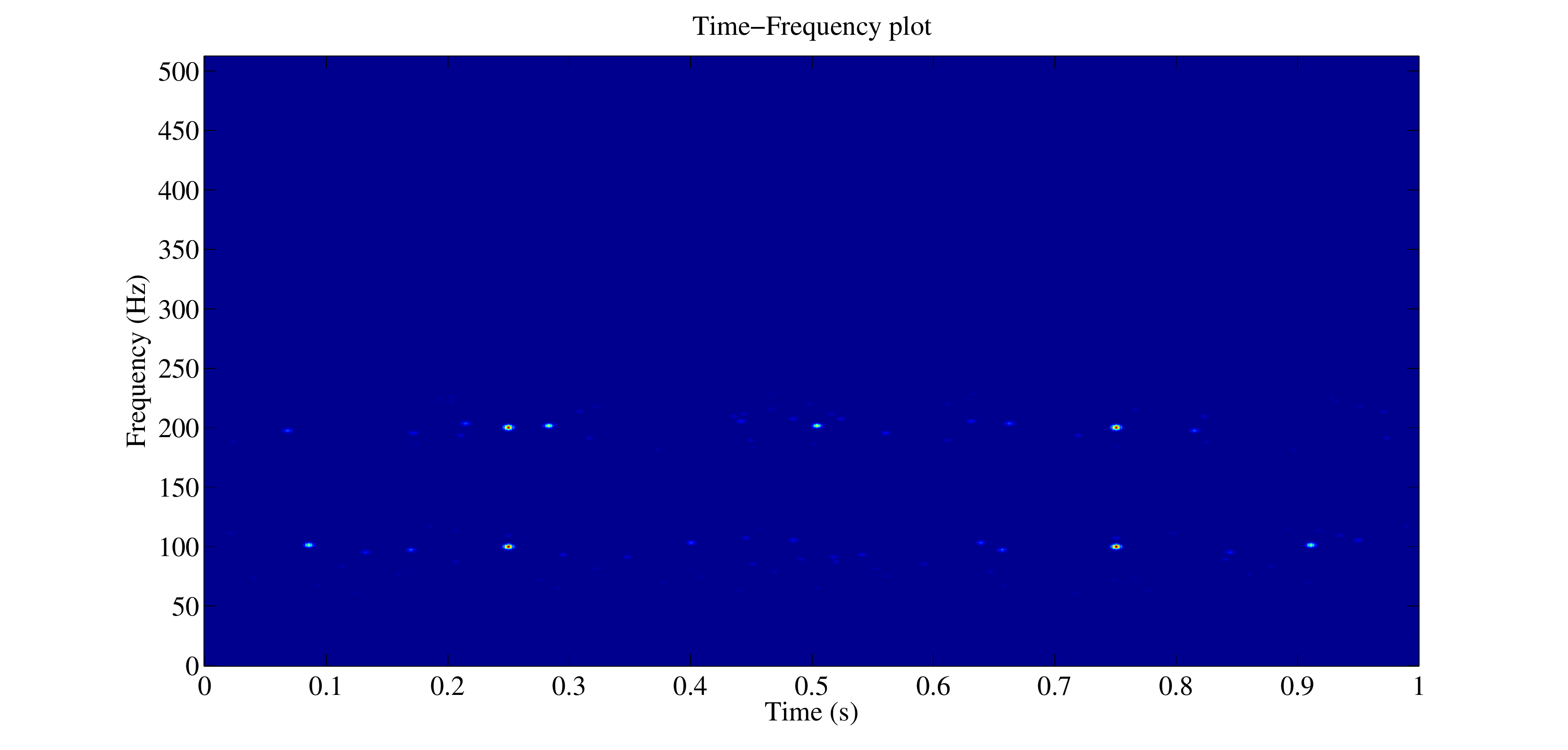}
\includegraphics[angle=0,width=0.5\textwidth,height=0.3\textwidth]{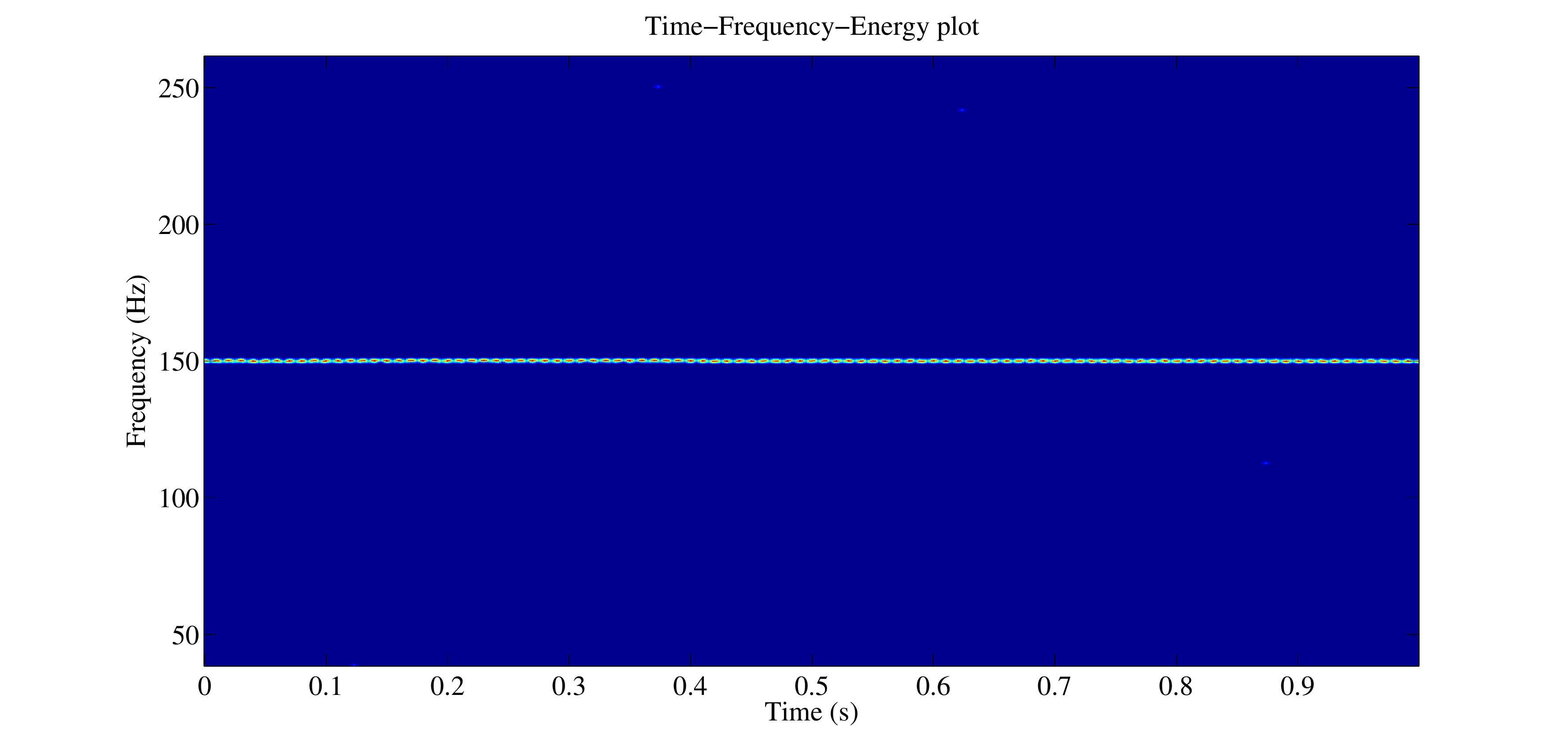}
\captionof{figure}{The TFE analysis of sum of two sinusoidal function $f_1=100$ Hz and $f_2=200$ Hz (with sampling frequency Fs = 1000 Hz, length $N = 1000$) using TFD-FT (upper) and TFD-IF (lower) without decomposition.}
\label{fig:Two_freq}
\end{figure}

Figure~\ref{fig:Two_freq} shows the TFE analysis of sum of sinusoidal functions of $f_1=100$ Hz and $f_2=200$ Hz frequency (with sampling frequency Fs = 1000 Hz, length $N = 1000$) using TFD-FT (upper) and TFD-IF (lower). Due to time averaging in the TFD-FT (upper) plot, it shows the signal under analysis is concentrated in two frequencies ($100$ Hz and $200$ Hz) and some time instants, whereas TFD-IF (lower) plot provides average frequency representation where signal is concentrated in average frequency ($f=150$ Hz), due to frequency averaging, and spread over all the time.

\begin{figure}[!t]
\centering
\includegraphics[angle=0,width=0.5\textwidth,height=0.3\textwidth]{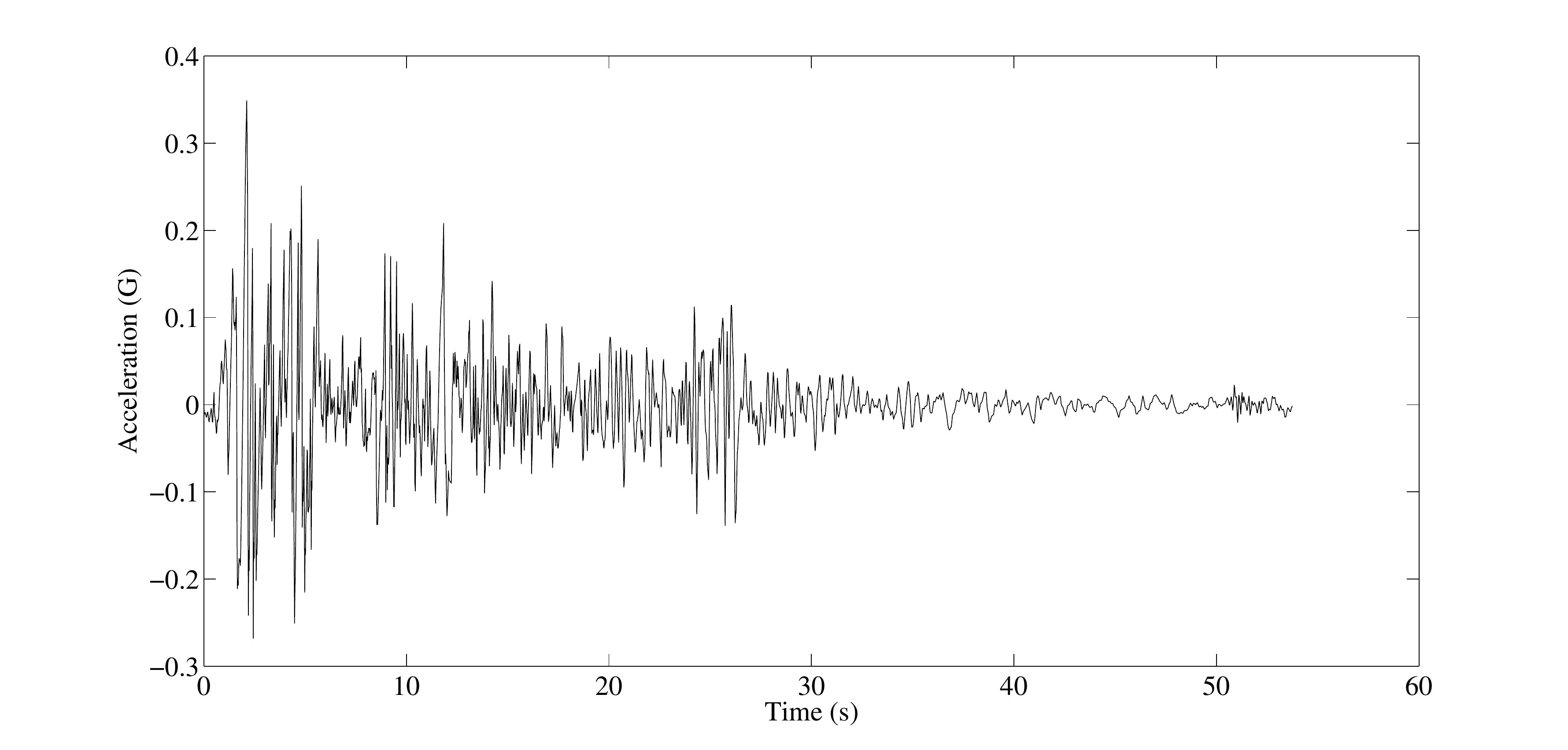}
\includegraphics[angle=0,width=0.5\textwidth,height=0.3\textwidth]{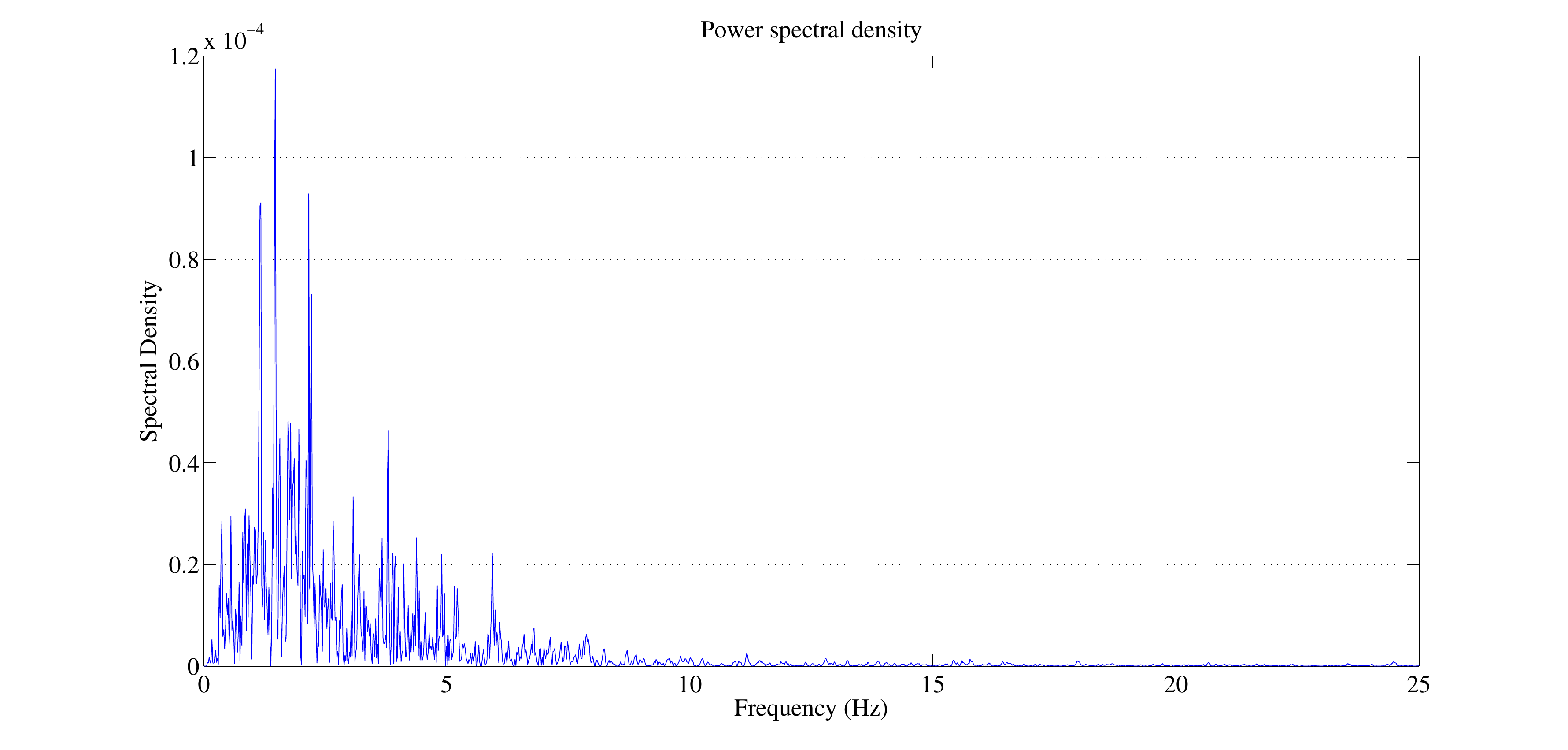}
\captionof{figure}{The Elcentro Earthquake May 18, 1940 North-South Component data (top), Fourier based power spectral density (PSD) (bottom).}
\label{fig:eqtfe}
\end{figure}

\begin{figure}[!t]
\centering
\includegraphics[angle=0,width=0.5\textwidth,height=0.3\textwidth]{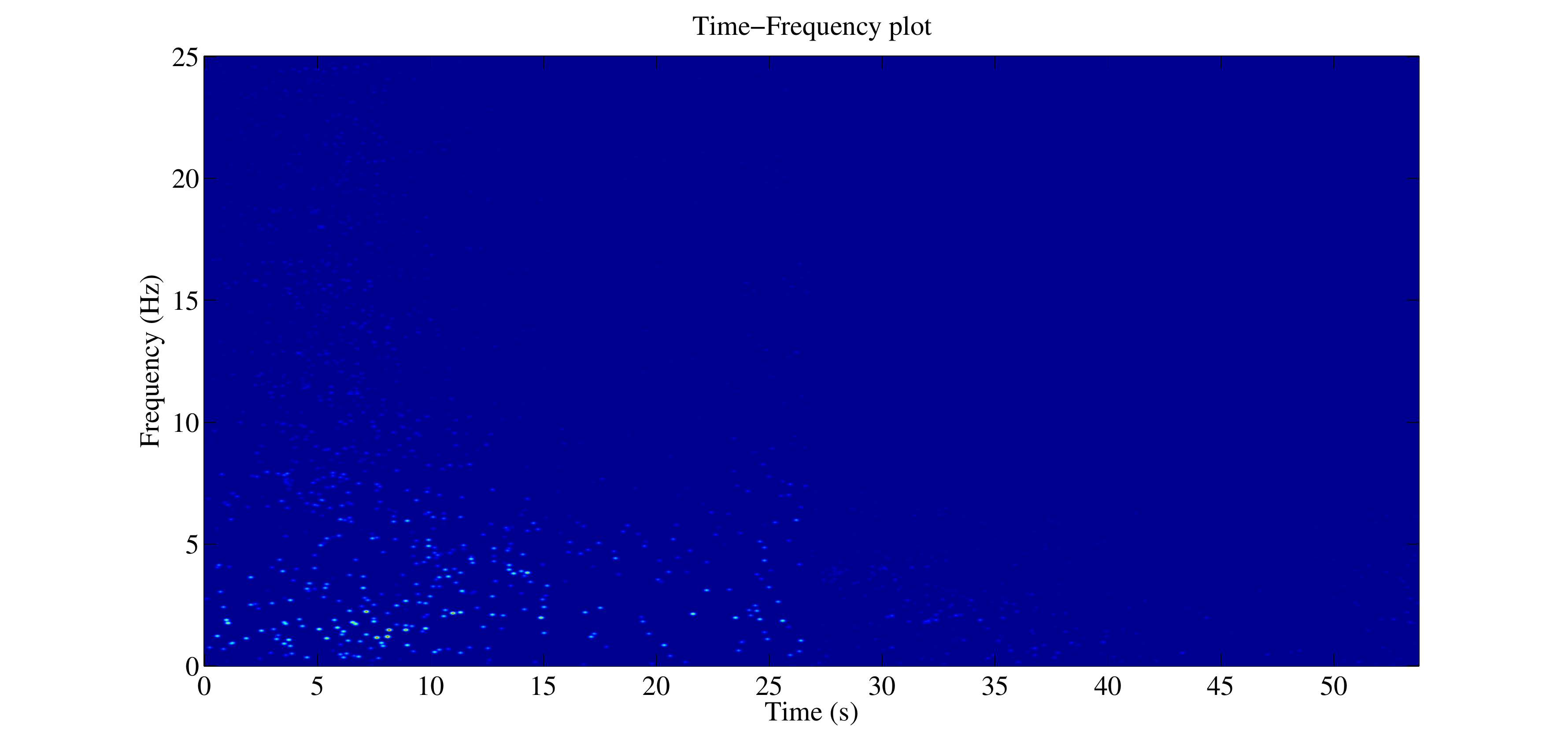}
\includegraphics[angle=0,width=0.5\textwidth,height=0.3\textwidth]{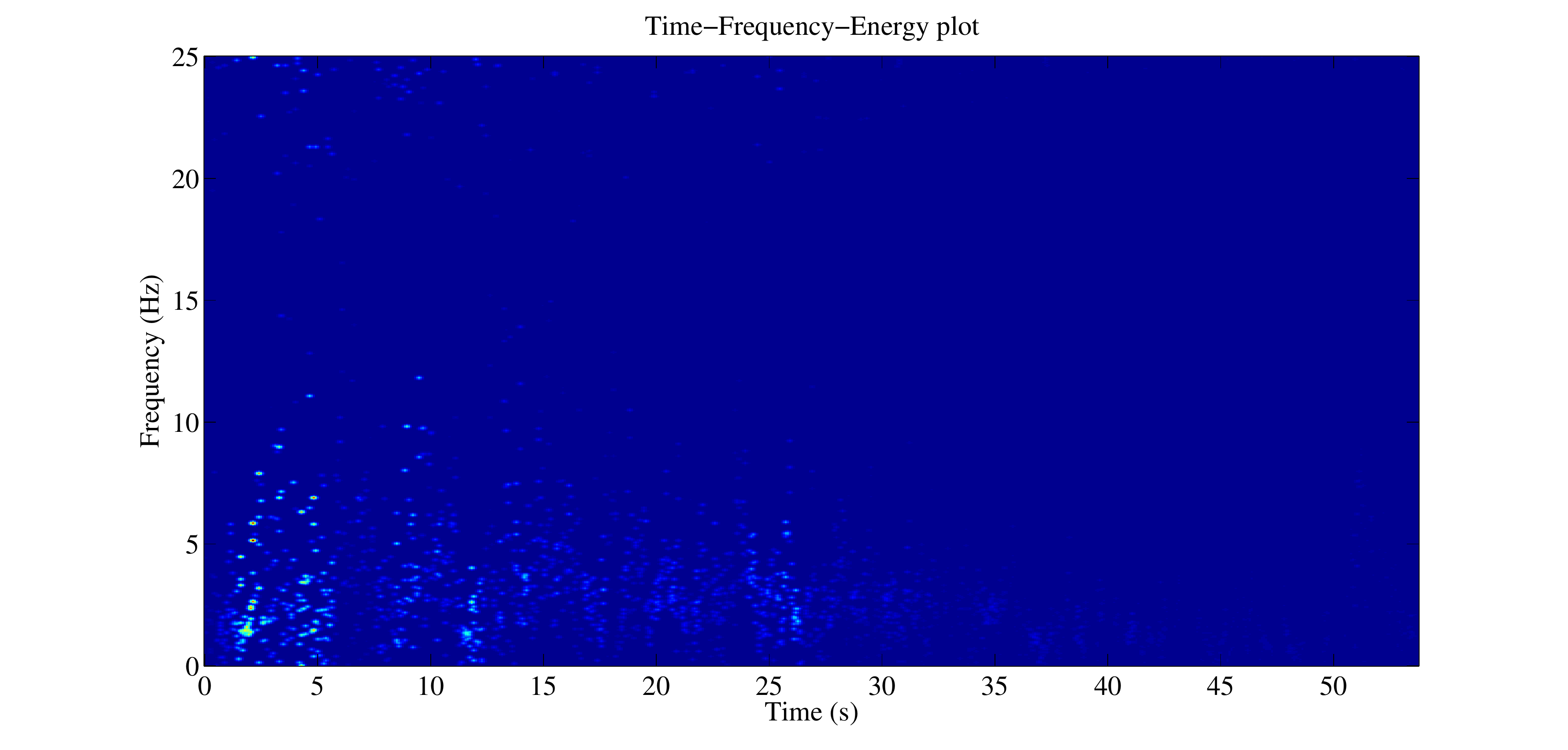}
\includegraphics[angle=0,width=0.5\textwidth,height=0.3\textwidth]{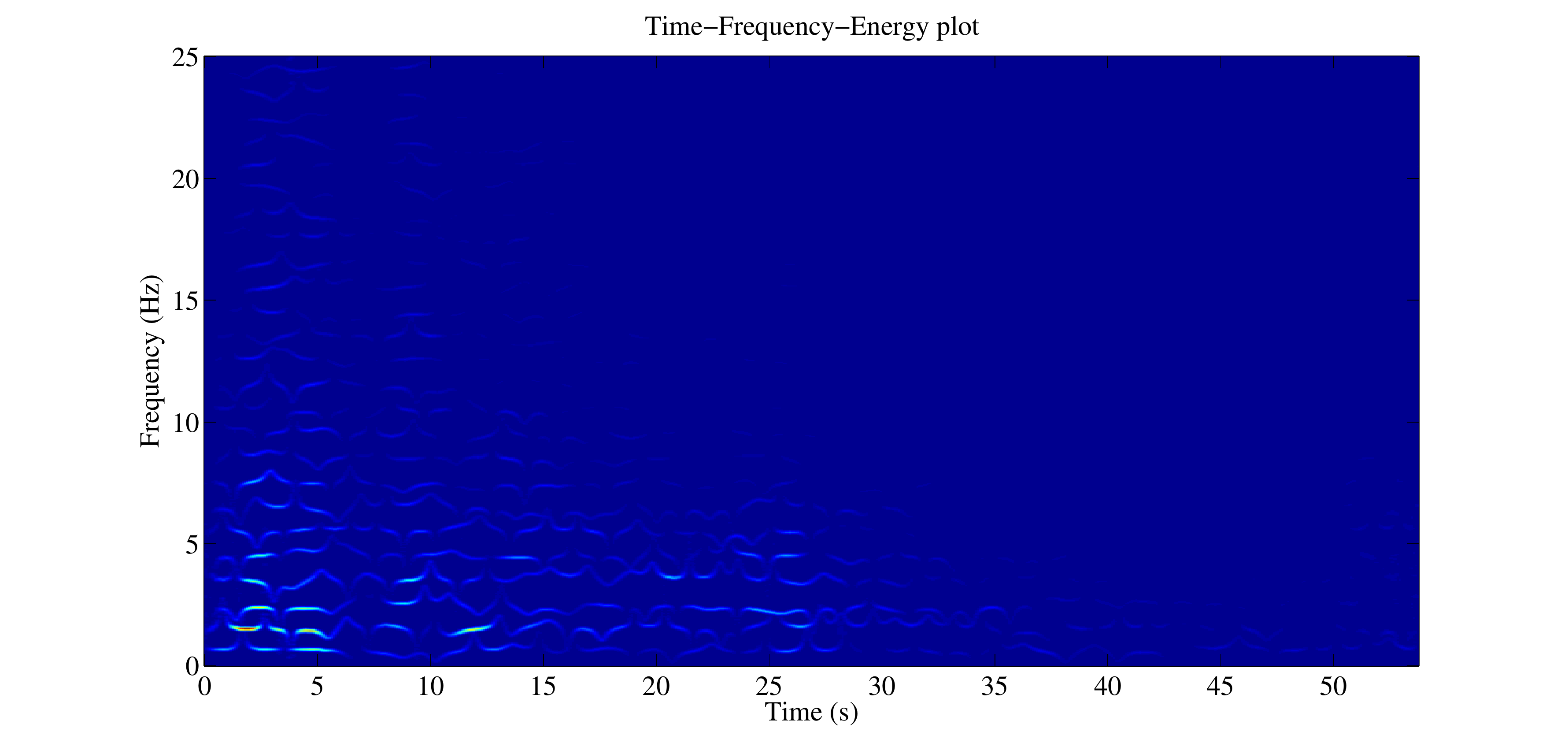}
\captionof{figure}{The TFE plot of the Elcentro Earthquake data (top to bottom) using the: (a) TFD-FT (b) TFD-IF and (c) TFD-IF with decomposition into 25 bands of 1 Hz each.}
\label{fig:eq_TFE1}
\end{figure}

\textbf{Example 4:} An Earthquake time series signal is a nonlinear and nonstationary data. The Elcentro Earthquake data (sampled at $F_s= 50Hz$) has been taken from~\cite{EQ33} and is shown in Figure~\ref{fig:eqtfe} (top one). The critical frequency range that matter in the structural design is less than $10Hz$, and the Fourier based power spectral density (PSD), Figure~\ref{fig:eqtfe} (bottom one), show that almost all the energy in this data is within $10Hz$.
The TFE distributions by the proposed method TFE-FT (top figure), using TFD-IF (middle figure) without decomposition and with decomposition into 25 bands of 1 Hz each are shown in Figure~\ref{fig:eq_TFE1}. These TFE distribution indicate that the maximum energy concentration is around $1.7Hz$ and 2 second. The TFE plot provide details of how the different waves arrive from the epical center to the recording station, e.g. the compression waves of small amplitude but higher frequency range of $10$ to $20Hz$, the shear and surface waves of strongest amplitude and lower frequency range of below $5Hz$ which does most of the damage, and other body shear waves which are present over the full duration of the data span.

\textbf{Discussion:} From Example 2 and Example 3, it is clear that if signal is concentrated in time then the proposed TFD-FT is performing better than TFD-IF, on the other hand, if signal is concentrated in frequency then performance of TFD-IF is better than TFD-FT. The propose TFD-FT contains true frequencies, i.e.  those frequencies which are present in the Fourier spectrum. Whereas, TFD-IF contains average frequencies present in signal. Thus, when we sum over the time then we obtain marginal spectrum which is true Fourier based PSD, which is one of the major advantage of the propose TFD-FT as compared to TFD-IF.

 \section*{CONCLUSION}
The instantaneous frequency (IF) is the time derivative of the instantaneous phase and it is an important parameter for the analysis of nonstationary signals and nonlinear systems. It is the basis of the time-frequency-energy (TFE) analysis of a signal via the inverse Fourier transform termed as Fourier-Hilbert spectrum (FHS). Dual to IF, we define the concept of `frequentaneous time' (FT) by the frequency derivative of phase which is the fundamental and important conceptual innovation of the this study. The frequentaneous time is the basis of the TFE analysis of a signal via the Fourier transform. The proposed TFD-FT contains only those frequencies which are present in the Fourier spectrum. The proposed frequentaneous time is valid for all types of signals such as monocomponent and multicomponent, narrowband and wideband, stationary and nonstationary, linear and nonlinear signals.

Simulations and numerical results demonstrate the efficacy, validity and superiority of the proposed `frequentaneous time' based TFD-FT for the TFE analysis of a signal via the Fourier transform as compared to IF and inverse Fourier transform based FHS method.
\section*{ACKNOWLEDGMENTS}
Author would like to show his gratitude to the Prof. SD Joshi (IITD), Prof. RK Pateny (IITD) and Dr. Kaushik Saha (Director, Samsung R\&D Institute India--Delhi) for sharing their wisdom and expertise with him during the course of this research.

\end{document}